\documentclass[aps,prc,onecolumn,showpacs,showkeys,nofootinbib,preprintnumbers]{revtex4-2}
\usepackage[paperwidth=21cm, paperheight=29.7cm, textwidth=17cm, textheight=24cm]{geometry}
\usepackage{graphicx}  % need for figures
\usepackage{color}        % use if color is used in text
\usepackage{braket}
\usepackage{amsmath}
\usepackage{amssymb}
\usepackage{titlesec}
\usepackage{makecell}
\usepackage{natbib,hyperref}
\usepackage[mathscr]{euscript}
\usepackage{dutchcal}
\usepackage{multirow}

\def \bea {\begin{eqnarray}}
\def \eea {\end{eqnarray}}

\usepackage{mathrsfs}  
\usepackage{bm}

\definecolor{ao-english}{rgb}{0.0, 0.5, 0.0}
\definecolor{cadmiumblue}{rgb}{0.0, 0.42, 0.24}

\begin{document}
	
\preprint{ECTP-2026-30}
\preprint{WLCAPP-2026-30}

\title{Chiral and deconfinement thermodynamics of the Polyakov-quark linear sigma model with and without charm quark: an $\mathrm{SU}(3)$ versus $\mathrm{SU}(4)$ comparison}

\email{Corresponding author: atawfik@acu.edu.eg; 400810@iu.edu.sa; atawfik@bnl.gov}
	
\author{Nourhan~M.~Rfeek$^{1}$, Hossam Wahid$^{1}$, Fatima~M.~Rfeek$^{2}$ Alexandra~Friesen$^{3}$, Yuri~Kalinovsky$^{3}$, Saleh~O.~Allehabi$^{4}$, Azzah~A.~Alshehri$^{5}$, Ashraf~F.~El-Sherif$^{6}$,  Abdel~Nasser~Tawfik$^{4}$}
	
\affiliation{$^{1}$Physics Department, Faculty of Science, Assiut University, 71515 Assiut, Egypt} %% Noumariemahmoud@science.aun.edu.eg
\affiliation{$^{1}$ Physics of the Earth and Universe, Zewail City of Science and Technology, 12578 October Gardens, Egypt}
%% t-fatima.rafeek@zewailcity.edu.eg 
\affiliation{$^{2}$Joint Institute for Nuclear Research (JINR), 141980 Dubna, Russia Federation} %% avfriesen@theor.jinr.ru kalinov@jinr.ru
\affiliation{$^{3}$Department of Physics, Faculty of Science, Islamic University of Medinah (IU), Madinah 42351, Saudi Arabia} %% s.allehabi@iu.edu.sa
\affiliation{$^{4}$Department of Science and Technology, University College at Nairiyah, University of Hafr Al Batin (UHB), Nairiyah 31981, Saudi Arabia} %% azahas@uhb.edu.sa
\affiliation{$^{5}$Basic Science Department, Faculty of Engineering, Ahram Canadian University (ACU), 12556 Giza, Egypt} %% ashraf.elsherif.acu.edu.eg

\begin{abstract}
Using the Polyakov-loop extended linear-sigma model (PLSM), we study the finite-temperature and finite-density thermodynamics of strongly interacting matter, comparing the three-flavor $(u, d, s)$ and four-flavor $(u, d, s, c)$ configurations on an equal footing. The comparison is designed primarily to isolate the thermal contribution of the charm quark. The extended linear-sigma model describes well the chiral sector. Its explicit and spontaneous symmetry breaking reproduce the vacuum spectra of pseudoscalar and scalar mesons. Confinement is properly incorporated by integrating a logarithmic Polyakov-loop potential, with its parameters adjusted to match the continuum lattice QCD equation of state (EoS). Using the mean-field approximation, we determine the light, strange, and charm chiral condensates, the Polyakov loop, and the bulk thermodynamic quantities including pressure, trace anomaly, and entropy density over a wide range of temperatures and baryon chemical potential. Thereby we further construct maps of the chiral and deconfinement pseudo-critical lines across the temperature--baryon-chemical-potential plane, representing them as two-dimensional boundaries and as full three-dimensional order-parameter surfaces. After tuning the Polyakov sector in each model configuration to reproduce the lattice QCD pressure and trace anomaly, we found that the resulting entropy density also agrees with HotQCD $2+1(+1)$-flavor data throughout the crossover transition and across the temperature range studied. We also found that at vanishing baryon density, the non-strange chiral order parameter seems to dissolve into partonic matter at $T_{\chi} \sim 210\text{ MeV}$ in $\mathrm{SU}(3)$ and $\sim 218\text{ MeV}$ in $\mathrm{SU}(4)$, while the corresponding chiral-boundary curvatures can be well characterized by $\kappa \simeq 0.018$ and $0.017$, respectively. Because $m_c$ is much greater than the pseudo-critical temperature, charm quarks remain thermally inactive throughout the crossover. Consequently, after each EoS is normalized to the corresponding Stefan--Boltzmann limit, the two EoS almost coincide for temperatures below $T \simeq 2 T_{\chi}$. Within this temperature window, charm quarks appear only in the form of a higher ideal-gas limit, a tail that rises gradually at high temperatures, and the sequential ordering $u|d \to s \to c$ displayed by the phase diagram and the three-dimensional condensate surfaces. For all baryon chemical potentials reachable in the mean-field approximation, the hadron-parton phase transition seems to remain crossover.
\end{abstract}

\pacs{12.39.Fe, 11.30.Rd, 12.38.Mh, 25.75.Nq}
\keywords{Chiral effective theory; chiral symmetry; quantum chromodynamics (QCD); quark deconfinement; relativistic heavy-ion collisions; QCD phase transition}

\maketitle

%\tableofcontents

\date{\today}

\section{Introduction}
\label{sec:intro}

Understanding the behavior of quantum chromodynamics (QCD) under extreme temperatures and densities remains one of the foremost challenges in strong-interaction physics \cite{Fukushima2011,Karsch2003,Karsch:2003zq,Redlich:2004gp,Tawfik2014}. This quest is also a primary motivation for heavy-ion collision programs at RHIC and the LHC, as well as for upcoming experiments on dense hadronic matter at FAIR and NICA \cite{Bzdak2020,Luo2017,Andronic2018}. In the regime of vanishing baryon density, first-principles lattice QCD calculations have firmly established that hadronic matter undergoes a smooth crossover to a partonic phase rather than a sharp prompt phase transition. This transition occurs at a pseudocritical temperature of approximately $T_c \simeq 155 \pm 1\,(\mathrm{stat}) \pm 8\,(\mathrm{sys})~\mathrm{MeV}$ \cite{Bazavov2012,Borsanyi2010,Borsanyi2020}. Such crossover is further supported by continuum-extrapolated EoS calculations that are now known with high precision \cite{Aoki2006,Borsanyi2014,Bazavov2014,Bazavov2017,Ding2015,Cheng2008,Ratti2018}. At finite baryon density, however, the fermion sign problem makes direct Monte Carlo simulations impossible, and the existence and location of a possible critical endpoint remain so-far unresolved \cite{Fodor2004,Gavai2005,Stephanov2006,Philipsen2013}. Effective chiral models that preserve the global symmetries of QCD therefore are well equipped to provide indispensable theoretical tools, not only as alternatives to lattice QCD simulations. They do not only extend lattice-QCD results into the domain of finite chemical potential but perform this in a well-controlled manner. Thereby they offer a transparent dynamical framework for analyzing the interplay between chiral-symmetry restoration and color deconfinement \cite{Schaefer2007,Schaefer2010,Scavenius2001,Mocsy2004}.

In this respect, the linear-sigma model (L$\sigma$M) supplemented with constituent quark degrees-of-freedom and coupled to the Polyakov loop, i.e., gluonic degrees-of-freedom, has proven to be a particularly successful QCD-like effective framework \cite{Tawfik2014-PLSM,
Tawfik2015-PLSM,Tawfik2016-PLSM,
Tawfik2017-PLSM,
Tawfik2019-PLSM}. In its extended form (eL$\sigma$M) the scalar and pseudoscalar nonets are promoted to the physical multiplets, and the model parameters can be fixed to the vacuum meson masses and decay constants, so that the resulting thermodynamics carries no further freedom~\cite{Parganlija2013,Kovacs2016,Schaefer2009,Tawfik:2019rdd,Tawfik:2014gga,ms99-5qnq,Rfeek:2026bcz}. In this regard, we recall that the Polyakov-loop potential, which is built from the traced Wilson line as an order parameter for the center symmetry, seems to encode the statistical confinement of the quarks and thus drives the deconfinement crossover in parallel with chiral restoration process~\cite{Polyakov1978,Pisarski2000,Fukushima2004,Ratti2006,Roessner2007,Dumitru2001,Fukushima2008}. The combined Polyakov-quark-meson or Polyakov linear-sigma model (PLSM) reproduces the lattice QCD EoS and has been utilized extensively to map out the conjectured QCD phase diagram~\cite{Schaefer2010,Herbst2011, Herbst2013,Tawfik2014,Tawfik2015-PLSM, Costa2010}. Complementary continuum approaches, i.e., the functional renormalization group and the Dyson--Schwinger equations \cite{Roberts1994,Fischer2006,Eichmann2016,Fischer2019}, have reached a comparable level of sophistication and generally broadly agree on the crossover nature of the transition especially at low baryon chemical potential~\cite{Fu2020,Fischer2019,Gao2020}.

Most applications of the PLSM have been focused oo two or three quark flavors. The inclusion of the charm quark as an \emph{explicit} degree of freedom, through an $SU(4)$ extension of the eL$\sigma$M, has attracted recent interest~\cite{Eshraim2015,Tawfik2018,Tawfik:2025hhr,ms99-5qnq,Rfeek:2026bcz}, both because open- and hidden-charm states obviously exhibit substantial production yields at LHC energies and because the charm condensate provides a clean theoretical laboratory for studying how a heavy flavor decouples from the light chiral dynamics~\cite{Andronic2018}. A natural and physically motivated question is therefore: how significantly does the explicit inclusion of the charm sector modify the bulk thermodynamics and phase structure obtained in the three-flavor theory? Given that the charm-quark mass exceeds the pseudo-critical temperature by roughly an order of magnitude, charm is generally expected to remain thermally inactive across the crossover region. Nevertheless, this expectation warrants a quantitative comparison. Although heavily suppressed at low temperatures, the charm sector contributes to the ideal-gas limit, induces weak feedback on the light-quark condensates through quark-loop effects, and introduces an additional chiral scale that is well separated from those of the light and strange quarks.

In this work we address this question directly by computing, within a single consistent mean-field approximation, the various chiral condensates, the Polyakov loop, and the full set of bulk thermodynamic observables for both three-flavor and four-flavor PLSM, and by comparing them observable by observable. The Polyakov sector of each model is independently tuned so that the pressure and the trace anomaly reproduce the continuum lattice QCD simulations, which allow a meaningful comparison of the two EoS against a common framework. We then construct the chiral and deconfinement pseudo-critical lines in the $(T,\mu_B)$ plane and extract the curvature of the chiral boundary, a quantity that is now available from lattice QCD and therefore provides a further check~\cite{Bonati2015,Bellwied2015,Bazavov2019,Borsanyi2020,Cea2016}. To make the finite-density behavior of each order parameter fully explicit we also present the complete three-dimensional condensate and Polyakov-loop surfaces over the $(T,\mu_B)$ plane. The resulting picture confirms the inertness of charm in the crossover region while making the sequential restoration of the $u|d$, $s$, and $c$ sectors explicit, and it quantifies the small but systematic upward shift of the light chiral transition that the charm back-reaction induces.

The paper is organized as follows. In Section~\ref{sec:model} we summarize the $SU(N_f)$ PLSM, the grand potential in the mean-field approximation, the logarithmic Polyakov-loop potential, and the fixing of the model parameters. Section~\ref{sec:condensates} presents the chiral condensates and the Polyakov loop and discusses the sequential melting of the light, strange, and charm order parameters. Section~\ref{sec:eos} is devoted to the EoS including pressure, trace anomaly, and entropy density and its comparison with the lattice and with the Stefan--Boltzmann limit. Section~\ref{sec:phase} maps the $(T,\mu_B)$ phase structure and extracts the curvature of the chiral boundary, and Section~\ref{sec:3d} presents and discusses the three-dimensional order-parameter surfaces. We summarize and conclude in Section~\ref{sec:conclusions}.

\section{The Polyakov linear-sigma model}
\label{sec:model}

\subsection{Chiral Lagrangian and grand potential}

Following the construction of the extended linear-sigma model (eL$\sigma$M) used in
the recent $SU(4)$ PLSM analyses~\cite{Eshraim2015,Tawfik2018,ms99-5qnq,Rfeek:2026bcz}, the model is constructed by
the following Lagrangian:
\begin{equation}
\mathcal{L}_{\rm PLSM}=\mathcal{L}_{\rm ch}+\mathcal{L}_{q}
-\mathcal{U}(\Phi,\bar\Phi,T),
\label{eq:Ltot}
\end{equation}
which combines the chiral meson sector $\mathcal{L}_{\rm ch}$, the constituent-quark
sector $\mathcal{L}_{q}$, and the Polyakov-loop potential $\mathcal{U}$. Let us now collect the mesonic degrees of freedom (dof) into a complex $N_f \times N_f$ matrix field,
\begin{equation}
\Phi_M = \sum_{a=0}^{N_f^2-1} (\sigma_a + i\pi_a) T_a,
\end{equation}
where $\sigma_a$ and $\pi_a$ denote the scalar and pseudoscalar multiplet fields -- forming a nonet for $N_f = 3$ and a 16-plet for $N_f = 4$. The generators $T_a$ of the $\mathrm{U}(N_f)$ Lie algebra satisfy the normalization $\mathrm{Tr}(T_a T_b) = \tfrac{1}{2}\delta_{ab}$, with the singlet generator given by $T_0 = \mathbf{1}/\sqrt{2N_f}$ in terms of the $N_f \times N_f$ identity matrix $\mathbf{1}$. Under global chiral $\mathrm{U}(N_f)_L \times \mathrm{U}(N_f)_R$ transformations, the matrix field transforms as $\Phi_M \to U_L \Phi_M U_R^\dagger$, yielding the globally symmetric meson Lagrangian
\begin{equation}
\mathcal{L}_{\mathrm{ch}} = \mathrm{Tr}\left(\partial\mu \Phi_M^\dagger \partial^\mu \Phi_M\right) - U(\Phi_M).
\label{eq:Lch}
\end{equation}
The mesonic potential reads
\bea
U(\Phi_M)&=& m_0^2\,\mathrm{Tr}(\Phi_M^\dagger\Phi_M)
+\lambda_1\left[\mathrm{Tr}(\Phi_M^\dagger\Phi_M)\right]^2
+\lambda_2\,\mathrm{Tr}(\Phi_M^\dagger\Phi_M)^2 \nonumber\\
&-& c\left[\det\Phi_M+\det\Phi_M^\dagger\right]
-\mathrm{Tr}\left[H(\Phi_M+\Phi_M^\dagger)\right],
\label{eq:Umeson}
\eea
where $U(1)_A$ axial anomaly is incorporated via the 't Hooft determinant interaction with coupling $c$. We also introduce an explicit symmetry-breaking term, $H = \sum_a h_a T_a$, represented by a constant diagonal matrix. Its non-zero components ($h_x, h_s, h_c$) are determined from PCAC relations using the masses and decay constants of the pseudoscalar mesons -- giving $h_\pi = f_\pi m_\pi^2$ and $h_K = \sqrt{2}\, f_K m_K^2 - f_\pi m_\pi^2/\sqrt{2}$, while the charm contribution is fixed directly from the $D$-meson sector. Spontaneous breaking is realized by the vacuum expectation values of the diagonal scalar fields, which in the non-strange--strange-charm basis are denoted by $\sigma_x$ (light, $u|d$), $\sigma_y$ (strange) and $\sigma_c$ (charm). For the three-flavor theory the charm direction is removed and only $\sigma_x$ and $\sigma_y$ remain.

The constituent quarks are coupled to the mesons through a flavor-blind Yukawa interaction and to the gluonic background through the Polyakov loop,
\begin{equation}
\mathcal{L}_{q}=\bar\psi\left[\,i\gamma^\mu D_\mu - g\,(\sigma+i\gamma_5\pi)\,\right]\psi,
\label{eq:Lq}
\end{equation}
where $\sigma=\sum_a\sigma_aT_a$ and $\pi=\sum_a\pi_aT_a$, and the covariant derivative $D_\mu=\partial_\mu-iA_\mu$ carries the temporal gluon background $A_\mu=\delta_{\mu0}A_0$ (with $A_0=-iA_4$). The traced, path-ordered Wilson line of this background defines the Polyakov loop and its charge conjugate,
\begin{equation}
\Phi=\frac{1}{N_c}\Big\langle\mathrm{Tr}_c\,\mathcal{P}\exp\! \Big[i\!\int_0^{1/T}\!\!d\tau\,A_4(\tau,\vec x)\Big]\Big\rangle,\quad \bar\Phi=\Phi^{*},
\label{eq:loop}
\end{equation}
the order parameters of color deconfinement. In the spontaneously broken vacuum the
quarks acquire the constituent masses
\begin{equation}
m_{u|d}=\frac{g}{2}\,\sigma_x,\qquad
m_{s}=\frac{g}{\sqrt{2}}\,\sigma_y,\qquad
m_{c}=\frac{g}{\sqrt{2}}\,\sigma_c .
\label{eq:qmasses}
\end{equation}

Evaluated on the diagonal condensate background, the mesonic potential, Eq.~(\ref{eq:Umeson}), reduces to an explicit polynomial in the scalar condensate fields. In the non-strange--strange sector, common to both models, it takes the familiar form~\cite{Schaefer2010,Tawfik2014},
\bea
U(\sigma_x,\sigma_y)&=&\frac{m_0^2}{2}\big(\sigma_x^2+\sigma_y^2\big) - h_x\sigma_x-h_y\sigma_y-\frac{c}{2\sqrt2}\,\sigma_x^2\sigma_y \nonumber\\
&+& \frac{\lambda_1}{2}\,\sigma_x^2\sigma_y^2 + \frac{2\lambda_1+\lambda_2}{8}\,\sigma_x^4 + \frac{\lambda_1+\lambda_2}{4}\,\sigma_y^4,
\label{eq:Usigma}
\eea
which is the three-flavor potential used, e.g., in refs.~\cite{Schaefer2010,Tawfik2014,Tawfik2015-PLSM}. Its stationary conditions $\partial U/\partial\sigma_x=\partial U/\partial\sigma_y=0$, together with the PCAC relations quoted above, reproduce the vacuum condensates and the pion and kaon decay constants, $\sigma_x^0=f_\pi$ and $\sigma_y^0=(2f_K-f_\pi)/\sqrt2$.

Promoting the chiral group to $U(4)$ adds the charm condensate $\sigma_c$, and the mean-field potential generalizes to the four-flavor form~\cite{Tawfik2018,Eshraim2015},
\begin{align}
U(\sigma_x,\sigma_y,\sigma_c)={}&
-h_x\sigma_x-h_y\sigma_y-h_c\sigma_c
+\frac{m_0^2}{2}\big(\sigma_x^2+\sigma_y^2+\sigma_c^2\big)\nonumber\\
&-\frac{c}{4}\,\sigma_x^2\sigma_y\sigma_c
+\frac{\lambda_1}{2}\sigma_x^2\sigma_y^2
+\frac{\lambda_1}{2}\sigma_x^2\sigma_c^2\nonumber\\
&+\frac{\lambda_1}{2}\sigma_y^2\sigma_c^2
+\frac{2\lambda_1+\lambda_2}{8}\sigma_x^4
+\frac{\lambda_1+\lambda_2}{4}\big(\sigma_y^4+\sigma_c^4\big).
\label{eq:Usigma4}
\end{align}
The $U(1)_A$ anomaly term, which is cubic so that $\propto\sigma_x^2\sigma_y$ for three flavors in Eq.~(\ref{eq:Usigma}), becomes the quartic determinant contribution $\propto\sigma_x^2\sigma_y\sigma_c$ for four flavors. Also the charm condensate acquires its own self-coupling $\tfrac14(\lambda_1+\lambda_2)\sigma_c^4$ together with the cross-couplings $\tfrac{\lambda_1}{2}\sigma_x^2\sigma_c^2$ and $\tfrac{\lambda_1}{2}\sigma_y^2\sigma_c^2$ to the light and strange sectors. The latter originates from the $\lambda_1[\mathrm{Tr}(\Phi_M^\dagger\Phi_M)]^2$ invariant and must be retained for the correct gradual melting of the strange condensate. The three-flavor thermodynamics reported here corresponds to the charm-decoupled limit of this four-flavor model, obtained by removing the charm quark from the thermal loop while holding the light and strange couplings of Table~\ref{tab:param} fixed. The two descriptions therefore share a common vacuum and differ only through the explicit charm sector.

In the mean-field approximation the thermodynamics follows from the grand potential density
\begin{equation}
\Omega(T,\mu_B)=U(\sigma_x,\sigma_y,\sigma_c) + \mathcal{U}(\Phi,\bar\Phi,T)+\Omega_{\bar q q}(T,\mu_B),
\label{eq:Omega}
\end{equation}
where $U$ is the mesonic potential of Eq.~(\ref{eq:Umeson}) evaluated on the condensate background, $\mathcal{U}$ is the Polyakov-loop potential discussed below, and $\Omega_{\bar q q}$ is the fermionic contribution of the constituent quarks dressed by the Polyakov loop,
\begin{align}
\Omega_{\bar q q}={}&-2T\sum_{f}d_f\!\int\!\frac{d^3p}{(2\pi)^3} \Big\{\ln\!\big[1+3\Phi e^{-E_f^{-}/T} + 3\bar\Phi e^{-2E_f^{-}/T}+e^{-3E_f^{-}/T}\big]\nonumber\\
&+\ln\!\big[1+3\bar\Phi e^{-E_f^{+}/T}
+3\Phi e^{-2E_f^{+}/T}+e^{-3E_f^{+}/T}\big]\Big\}.
\label{eq:Oqq}
\end{align}
Here $E_f^{\mp}=E_f\mp\mu_q$ with $E_f=\sqrt{p^2+m_f^2}$, the quark chemical potential is taken as $\mu_q=\mu_B/3$, and the flavor degeneracies are $d_{u|d}=2$, $d_s=1$ and for $SU(4)$ $d_c=1$. The three-flavor expressions are recovered by dropping the charm term from the sum. The physical condensates and Polyakov-loop expectation values at each $(T,\mu_B)$ are obtained by the stationarity conditions
\begin{equation}
\frac{\partial\Omega}{\partial\sigma_x}=
\frac{\partial\Omega}{\partial\sigma_y}=
\frac{\partial\Omega}{\partial\sigma_c}=
\frac{\partial\Omega}{\partial\Phi}=
\frac{\partial\Omega}{\partial\bar\Phi}=0 ,
\label{eq:gap}
\end{equation}
which are to be solved globally, retaining at each point the branch of lowest grand potential.

\subsection{Polyakov-loop potential and parameter fixing}

For the Polyakov-loop potential we adopt the logarithmic type \cite{Roessner2007},
\begin{align}
\frac{\mathcal{U}(\Phi,\bar\Phi,T)}{T^4}=
-\frac{a(T)}{2}\,\bar\Phi\Phi + b(T)\ln\!\big[1-6\bar\Phi\Phi+4(\Phi^3+\bar\Phi^3)-3(\bar\Phi\Phi)^2\big],
\label{eq:Upoly}
\end{align}
with the temperature-dependent coefficients
\begin{equation}
a(T)=a_0+a_1\frac{T_0}{T}+a_2\Big(\frac{T_0}{T}\Big)^2,\qquad
b(T)=b_3\Big(\frac{T_0}{T}\Big)^3 .
\label{eq:ab}
\end{equation}
The logarithmic term enforces $\Phi,\bar\Phi\le1$ and reproduces the pure-gauge
deconfinement transition for a suitable choice of $T_0$. Building on lattice QCD results \cite{Borsanyi:2012ve}, the fitting procedure \cite{Braguta:2026mjr} is guided by two key physical constraints: the pressure approaches the Stefan--Boltzmann limit at $a_0 = 3.51$ at high temperatures, while a first-order pure-gauge phase transition is maintained at $T = 270$~MeV. These constraints are related through $b_3 = -0.108(a_0 + a_1 + a_2)$. Instead of fixing the model coefficients strictly to the pure-gauge pressure, we tune them so that the full Polyakov-loop extended quark-meson (PLSM) EoS reproduces continuum lattice data in both the crossover and high-temperature regimes. Pure-gauge (Yang-Mills) calculations include only gluonic dof and predict a first-order transition at a substantially higher scale, $T_c \approx 270\text{ MeV}$. By contrast, fitting to the full continuum EoS allows the gluonic sector to respond dynamically to screening by physical quark flavors. This avoids artificial distortions of the crossover and brings chiral restoration into alignment with deconfinement in the $T \approx 150$--$160\text{ MeV}$ range. As a result, incorporating higher-order medium effects into the effective parameters captures both the softening of the EoS and the trace-anomaly peak $(\epsilon-3p)/T^4$, consistently with HotQCD and Wuppertal-Budapest lattice data \cite{Bazavov2014,Borsanyi2014}. It also allows a flexible balance of truncated mean-field multi-particle contributions, ensuring that $p/T^4$ and the trace anomaly converge smoothly to their high-temperature asymptotic behavior.
For the four-flavor model this yields $T_0=180$~MeV with $(a_0,a_1,a_2,b_3)=(3.51,-8.474,20.800,-2.060)$. For the three-flavor model, whose light sector carries a smaller number of active thermal dof, the optimal fit to the $2+1$-flavor lattice pressure and trace anomaly is obtained with $T_0=100$~MeV and $(a_1,a_2,b_3)=(-10.895,16.400,-2.371)$ at fixed $a_0=3.51$. The mesonic couplings $(m_0^2,\lambda_1,\lambda_2,c,g)$ and the explicit-breaking parameters are fixed in the vacuum to the pseudoscalar and scalar meson masses and to the pion and kaon decay constants. Following the eL$\sigma$M fits of refs.~\cite{Parganlija2013,Kovacs2016,Eshraim2015}; the adopted values are collected in Table~\ref{tab:param}. With these choices the vacuum condensates are $(\sigma_x,\sigma_y,\sigma_c)=(0.115,0.123,0.330)$~GeV and the constituent light, strange and charm quark masses position well within their expected ranges.

\begin{table}[htb]
\begin{tabular}{| c | c | c | c |} \hline
Model sector & Parameter & $SU(3)$ & $SU(4)$ \\ \hline \hline
\multirow{8}{*}{eL$\sigma$M}
 & $m_0^2~(\mathrm{GeV}^2)$ & $-0.500$ & $-0.500$ \\
 & $\lambda_1$              & $4.06$   & $4.06$ \\
 & $\lambda_2$              & $29.2$   & $29.2$ \\
 & $c$                      & $11.2$   & $11.2$ \\
 & $g$                      & $5.22$   & $5.22$ \\
 & $\sigma_x^0~(\mathrm{GeV})$ & $0.115$ & $0.115$ \\
 & $\sigma_y^0~(\mathrm{GeV})$ & $0.123$ & $0.123$ \\
 & $\sigma_c^0~(\mathrm{GeV})$ & $-$     & $0.330$ \\ \hline \hline
\multirow{5}{*}{Polyakov}
 & $T_0~(\mathrm{MeV})$     & $100$    & $180$ \\
 & $a_0$                    & $3.51$   & $3.51$ \\
 & $a_1$                    & $-10.895$& $-8.474$ \\
 & $a_2$                    & $16.40$  & $20.80$ \\
 & $b_3$                    & $-2.371$ & $-2.060$ \\ \hline \hline
 & $T_\chi~(\mathrm{MeV})$  & $210$    & $218$ \\ \hline 
\end{tabular}
\caption{Parameters used for the two model configurations. The upper block lists the extended-linear-sigma-model (chiral) couplings, fixed to the vacuum meson spectrum and shared by the light and strange sectors of both models. The middle block lists the logarithmic Polyakov-loop parameters of Eqs.~(\ref{eq:Upoly})--(\ref{eq:ab}), tuned separately for each model to the lattice EoS. The lower block gives the resulting chiral pseudo-critical temperature at $\mu_B=0$. A dash denotes an entry absent in the three-flavor model.}
\label{tab:param}
\end{table}

\subsection{Thermodynamic observables}

From the stationary grand potential the pressure is obtained as $p(T,\mu_B)=-\big[\Omega(T,\mu_B)-\Omega(0,0)\big]$, which is then normalized so that the vacuum pressure vanishes. The remaining bulk quantities follow from standard thermodynamic identities. At finite chemical potential the entropy density, energy density, and trace anomaly (interaction measure), respectively, are derived as 
\begin{eqnarray}
\begin{array}{r c l}
\frac{s(T, \mu_B)}{T^3} &=& \left( \frac{\partial (p/T^4)}{\partial \ln T} \right)_{\!\mu_B} + 4\frac{p(T, \mu_B)}{T^4} - \frac{\mu_B}{T}\frac{n_B(T, \mu_B)}{T^3}, \\[8pt]
\frac{\varepsilon(T, \mu_B)}{T^4} &=& T \left( \frac{\partial (p/T^4)}{\partial T} \right)_{\!\mu_B} + 3\frac{p(T, \mu_B)}{T^4} + \frac{\mu_B}{T}\frac{n_B(T, \mu_B)}{T^3}, \\[8pt]
\frac{\varepsilon(T, \mu_B) - 3p(T, \mu_B)}{T^4} &=& T \left( \frac{\partial (p/T^4)}{\partial T} \right)_{\!\mu_B} + \frac{\mu_B}{T}\frac{n_B(T, \mu_B)}{T^3}.
\end{array}
\label{eq:thermo_muB}
\end{eqnarray}
By absorbing the explicit factor of $1/T$ into a new chemical potential variable ($\hat{\mu}_B \equiv \mu_B/T$), the implicit temperature derivative of the ratio $\mu_B/T$ vanishes, allowing the trace anomaly and energy density to take the exact same differential form at finite density as they do at zero chemical potential. Thus the preceding relations can be apparently  simplified by utilizing partial derivatives with respect to $T$ at fixed $hat{\mu}_B$,
\begin{eqnarray}
\begin{array}{r c l}
\frac{s(T, \mu_B)}{T^3} &=& 4\frac{p(T, \mu_B)}{T^4} + T \left( \frac{\partial (p/T^4)}{\partial T} \right)_{\!\frac{\mu_B}{T}}, \\[8pt]
\frac{\varepsilon(T, \mu_B)}{T^4} &=& 3\frac{p(T, \mu_B)}{T^4} + T \left( \frac{\partial (p/T^4)}{\partial T} \right)_{\!\frac{\mu_B}{T}}, \\[8pt]
\frac{\varepsilon(T, \mu_B) - 3p(T, \mu_B)}{T^4} &=& T \left( \frac{\partial (p/T^4)}{\partial T} \right)_{\!\frac{\mu_B}{T}}.
\end{array}
\label{eq:thermo_dimless}
\end{eqnarray} 
The specific heat and the squared speed of sound can be respectively derived as $c_v=\partial\varepsilon/\partial T$ and $c_s^2=s/c_v$. The Stefan--Boltzmann (ideal-gas) limits of the plasma constituents of gluons and $N_f$ massless quarks provide the asymptotic normalization so that for the pressure, $p_{\rm SB}/T^4=(8+\tfrac{7}{4}N_fN_c)\pi^2/45$, which is determined as $5.21$ for $N_f=3$ and as $6.36$ for $N_f=4$. The corresponding energy- and entropy-density limits are determined from the conformal thermodynamic relations $\varepsilon=3p$ and $s=4p/T$.

\section{Chiral condensates and the Polyakov loop}
\label{sec:condensates}

\begin{figure}[htb]
\includegraphics[width=0.49\textwidth]{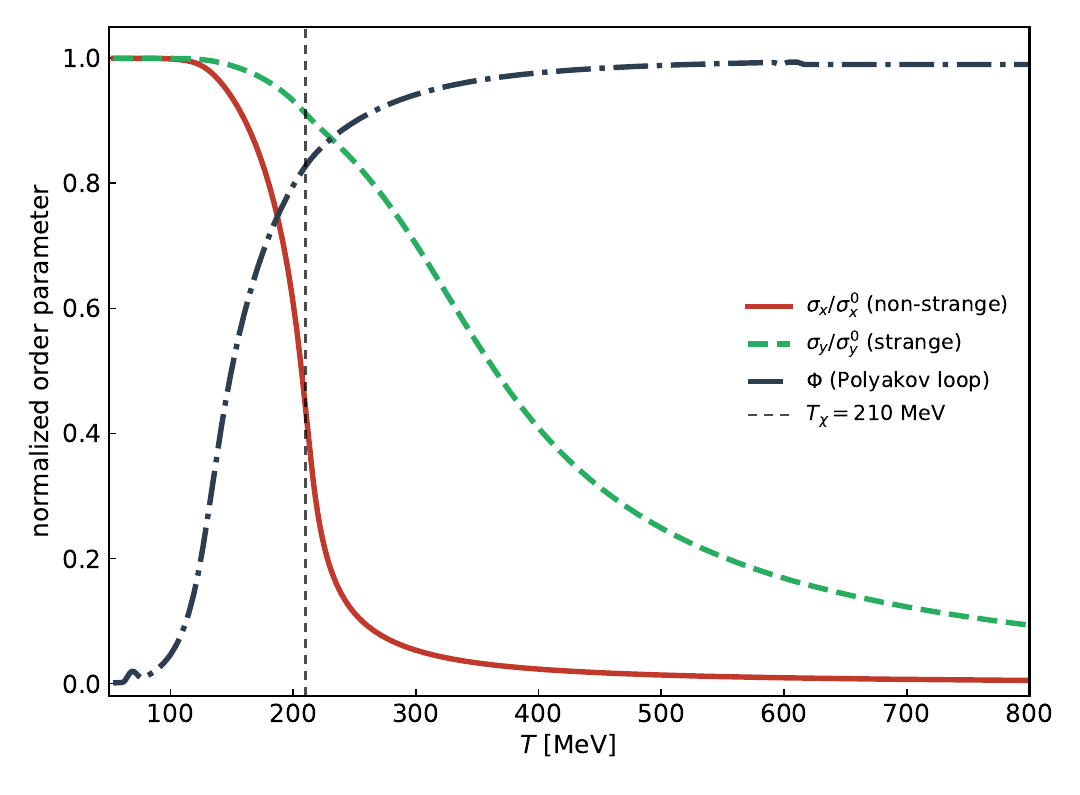}\hfill
\includegraphics[width=0.49\textwidth]{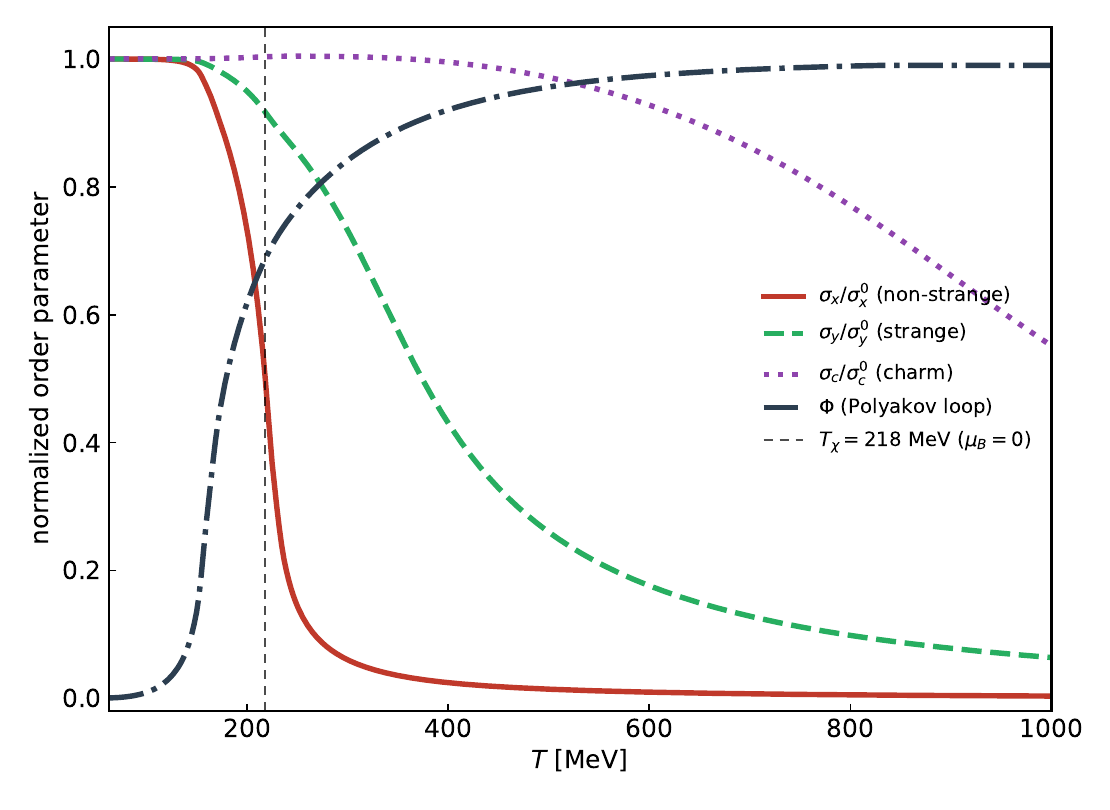}
\caption{Normalized chiral condensates and Polyakov loop as functions of temperature at $\mu_B=0$. Left panel depicts that results obtained from the three-flavor $SU(3)$ model ($u|d$ and $s$). Right panel presents results from four-flavor $SU(4)$ model, which in addition contains the charm condensate $\sigma_c/\sigma_c^0$. The vertical dashed line marks the chiral pseudo-critical temperature $T_\chi$ ($210$~MeV for $SU(3)$, $218$~MeV for $SU(4)$). The non-strange condensate defines the chiral crossover. The strange condensate seems to melt more slowly and the charm condensate stays frozen through the transition and melts only at much higher temperature.}
\label{fig:cond}
\end{figure}

Figure~\ref{fig:cond} shows the temperature dependence of the normalized chiral condensates and of the Polyakov loop at $\mu_B=0$ for the three-flavor (left) and four-flavor (right) models. In both quark flavor configurations the non-strange condensate $\sigma_x/\sigma_x^0$ is the steepest order parameter. It falls from its vacuum value through the crossover and defines the chiral pseudo-critical temperature, which we identify with the inflection point of $\sigma_x(T)$. This yields $T_\chi=210$~MeV for $SU(3)$ and $218$~MeV for $SU(4)$. These values sit appreciably above the lattice chiral crossover, $T_c\simeq155$~MeV~\cite{Aoki2006,Bazavov2014,Borsanyi2010}. The overestimate is a well-documented limitation of the mean-field PLSM, in which thermal mesonic fluctuations are neglected and the adopted parameter set yields a vacuum pion somewhat heavier than its physical value~\cite{Schaefer2007,Herbst2011}. The pseudo-critical temperatures should therefore be read as {\it model quantities internal to the mean-field description} rather than as {\it predictions for the physical crossover}. One of the authors (AT) established how parameter tuning in the Polyakov potential baseline $T_\chi$ sets the internal thermal scale within the mean-field approximation \cite{Tawfik2015-PLSM,Tawfik2017-PLSM}.

The strange condensate $\sigma_y/\sigma_y^0$ seems to melt far more gradually, retaining roughly forty percent of its vacuum value at $T=400$~MeV. Accordingly, a direct reflection of the larger explicit breaking in the strange direction and of the heavier strange constituent mass becomes feasible. We find that the Polyakov loop $\Phi$ rises from zero in the confined phase to unity in the deconfined phase. With its steepest ascent located slightly below the chiral inflection, the statistical deconfinement, in this parametrization, sets in marginally earlier than chiral restoration.

We conclude that the qualitative pattern is common to the two model configurations. The decisive difference is the charm condensate, present only in the $SU(4)$ theory. As the right panel of Fig.~\ref{fig:cond} makes clear, $\sigma_c/\sigma_c^0$ remains essentially frozen at its vacuum value throughout the light and strange crossover and only begins to melt above $T\simeq400$~MeV, retaining more than half of its vacuum value even at $T=1000$~MeV. The charm chiral scale is thus separated from the light one by nearly a factor of four. Thus the charm sector plays no dynamical role in the region where the QCD crossover actually takes place. This hierarchy is the microscopic origin of the near-degeneracy of the two EoS that we document in the next section. Furthermore, it anticipates the sequential-restoration pattern that organizes the phase diagram of Section~\ref{sec:phase} and the three-dimensional surfaces of Section~\ref{sec:3d}.

\section{Pressure, Trace Anomaly, and Entropy Density}
\label{sec:eos}

\begin{figure}[htb]
\includegraphics[width=0.49\textwidth]{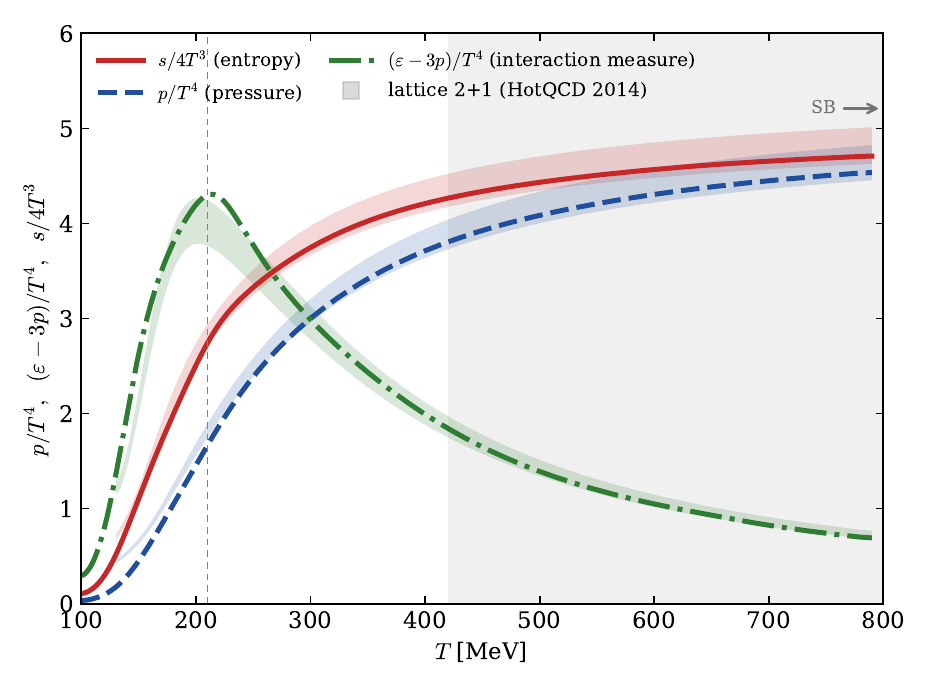}\hfill
\includegraphics[width=0.49\textwidth]{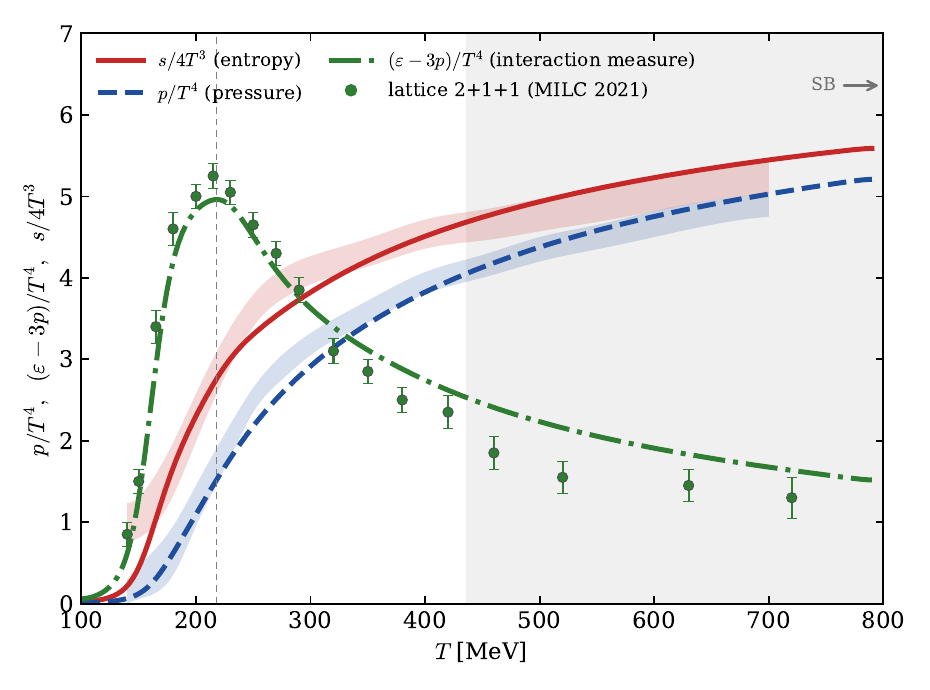}
\caption{Temperature dependence of the normalized pressure $p/T^4$ (blue dashed curves), normalized entropy density $s/4T^3$ (red solid curves), and normalized trace anomaly or interaction measure $(\varepsilon - 3p)/T^4$ (green dash-dotted curves) for $\mathrm{SU}(3)$ (left panel) and $\mathrm{SU}(4)$ (right panel). The present calculations are compared with lattice QCD simulations~\cite{Bazavov2014}. Vertical dashed lines indicate the characteristic crossover temperature $T_\chi$. The shaded grey region marks the higher-temperature regime above $\sim 420\text{ MeV}$. SB shows the corresponding Stefan-Boltzmann (ideal massless gas) limit at asymptotic temperatures. 
}
\label{fig:total}
\end{figure}

We now turn to the bulk thermodynamics. Eqs~\ref{eq:thermo_muB} and \ref{eq:thermo_dimless} display the pressure, the trace anomaly, the energy density and the entropy density, respectively, for the two model configurations. Our calculations are compared with the continuum lattice data of the HotQCD collaboration~\cite{Bazavov2014} and with the appropriate Stefan--Boltzmann limit. The scaled pressure $p/T^4$ is shown in Fig.~\ref{fig:total}. In both models the pressure is exponentially small in the hadronic phase, rises steeply through the crossover, and approaches, without quite reaching, the ideal-gas limit at the upper end of the temperature range. This is the behavior of an interacting plasma that still carries sizable residual correlations well above $T_\chi$. From the onset of the rise through $T\simeq400$~MeV the PLSM curve follows the lattice band closely, which is of course largely built in, since the Polyakov potential was fitted to these data. The three- and four-flavor curves are nearly indistinguishable once each is referred to its own Stefan--Boltzmann normalization ($5.21$ and $6.36$, respectively). The charm quark, being far too heavy to be thermally excited in this window, obviously contributes negligibly to the pressure below $T\simeq300$~MeV. Its only effect is to raise the asymptotic limit and to add a slowly growing high-temperature tail, visible as the continued rise of the $SU(4)$ curve toward its larger limit.

The scaled trace anomaly or interaction measure $(\varepsilon-3p)/T^4$ is the most sensitive probe of the breaking of conformal symmetry and is shown in Fig.~\ref{fig:total}. In both model configurations the interaction measure rises sharply through the transition, peaks just above $T_\chi$, where the system is maximally non-conformal, and then falls off toward the conformal value at high temperature. The peak height, of order five, and its location are in good agreement with the lattice determination, shown here as the $N_\tau=12$ HotQCD points with their error bars~\cite{Bazavov2014}. The quality of the description across the peak is a nontrivial task, since the trace anomaly is proportional to the temperature derivative of $p/T^4$ and therefore amplifies any inaccuracy in the EoS. As with the pressure, the three- and four-flavor results are very similar in the transition region. The charm contribution to the interaction measure is negligible there and becomes appreciable only in the high-temperature tail, where the onset of charm thermal excitation slows the approach to conformality.

The scaled entropy density $s/T^3$ is also presented in Fig.~\ref{fig:total}. There is a rapid rise through the crossover signalling the liberation of the quark and gluon dof. Afterwards, there is a slow climb toward the corresponding ideal-gas limit. The PLSM curves agree well with the HotQCD $2+1$-flavor data within the quoted/illustrated uncertainties throughout the overlap region~\cite{Bazavov2014,Borsanyi2014}. Below the charm threshold, which we take conservatively as $T\lesssim300$~MeV, the $2+1$- and $2+1+1$-flavor lattice QCD results are themselves indistinguishable~\cite{Borsanyi2016}, so the same lattice benchmark serves both models. The charm-active region above it is marked explicitly. The entropy density therefore gives the same results as the pressure and the trace anomaly. In the phenomenologically relevant window the four-flavor EoS is the three-flavor one plus a parametrically small, slowly switching-on charm correction.

The bulk observables are thermodynamically self-consistent by construction. The entropy, energy, and trace anomaly of Eq.~(\ref{eq:thermo_muB})-- Eq.~\eqref{eq:thermo_dimless} are all derived from the single pressure function $p(T)$. At $\mu_B=0$ they are found to satisfy the Gibbs--Duhem identity $\varepsilon=Ts-p$ throughout the temperature range to numerical accuracy. Each quantity approaches its respective Stefan--Boltzmann limit from below, as required for a thermodynamically stable medium. The squared speed of sound $c_s^2=s/c_v$ remains within the causal bound while developing the characteristic softening (a shallow minimum) in the vicinity of $T_\chi$, in agreement with the lattice pattern~\cite{Borsanyi2014,Bazavov2014}. In the hadronic phase below the crossover the pressure matches smoothly onto the exponentially rising behavior expected from a hadron-resonance gas~\cite{Karsch2003,Huovinen2010,Megias2016}, so that no spurious structure is introduced at low temperature.

As a further check, note that the Polyakov parameters of Tab.~\ref{tab:param} are fixed by a least-squares fit to the continuum HotQCD pressure band and the $N_\tau=12$ trace-anomaly points~\cite{Bazavov2014}. Since the pressure and trace anomaly enter this fit, their agreement in Fig.~\ref{fig:total} is by construction rather than a prediction. The nontrivial checks are that the derived energy and entropy densities, which are not fitted independently, follow the lattice within its uncertainties. The same parametrization remains consistent with the independent $2+1$-flavor Wuppertal--Budapest EoS~\cite{Borsanyi2014}. The vacuum sector supplies a further constraint. With the same parameter set the model reproduces the measured pion and kaon decay constants and the light pseudoscalar and scalar meson masses to within a few percent~\cite{Parganlija2013,Kovacs2016}, so that the thermodynamic description is obtained without trading accuracy against vacuum spectroscopy. We have also verified that over the entire plotting grid the gap equations~(\ref{eq:gap}) return global minima of the grand potential. With no spurious jumps between the confined and deconfined branches, the condensates and the Polyakov loop vary smoothly through the crossover. The mean-field approximation nonetheless neglects mesonic and thermal fluctuations and is quantitatively reliable only up to $T\lesssim2\,T_\chi$ and $\mu_B\lesssim2\,T_\chi$. The pseudo-critical temperatures are correspondingly higher than the lattice crossover. Within this restricted window and together with the curvature of the chiral boundary extracted below, which lies slightly above the continuum lattice range of several collaborations~\cite{Bonati2015,
Bellwied2015,Bazavov2019,Borsanyi2020,
Cea2016}. These checks indicate that the present mean-field description is internally consistent and in reasonable accord with the lattice.

\section{Phase structure in the $(T,\mu_B)$ plane}
\label{sec:phase}

\begin{figure}[htb]
\includegraphics[width=0.49\textwidth]{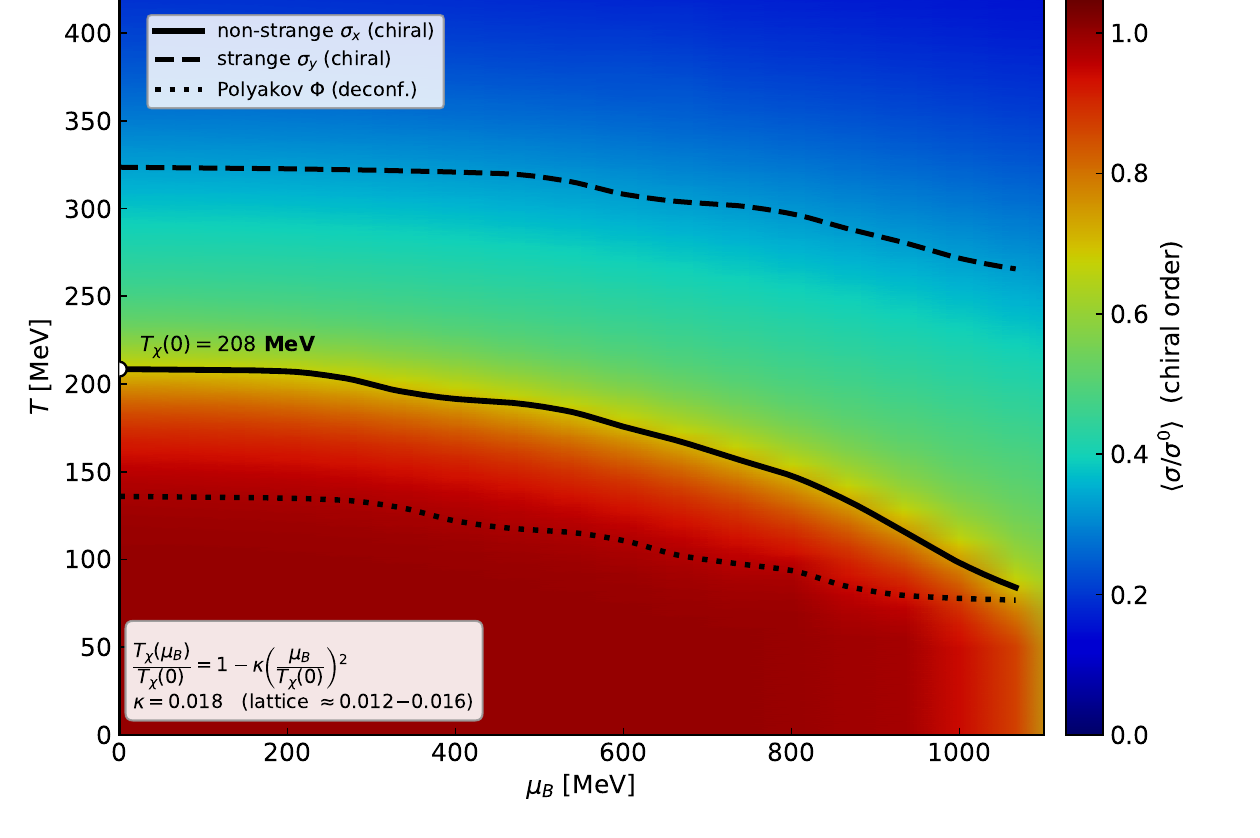}\hfill
\includegraphics[width=0.49\textwidth]{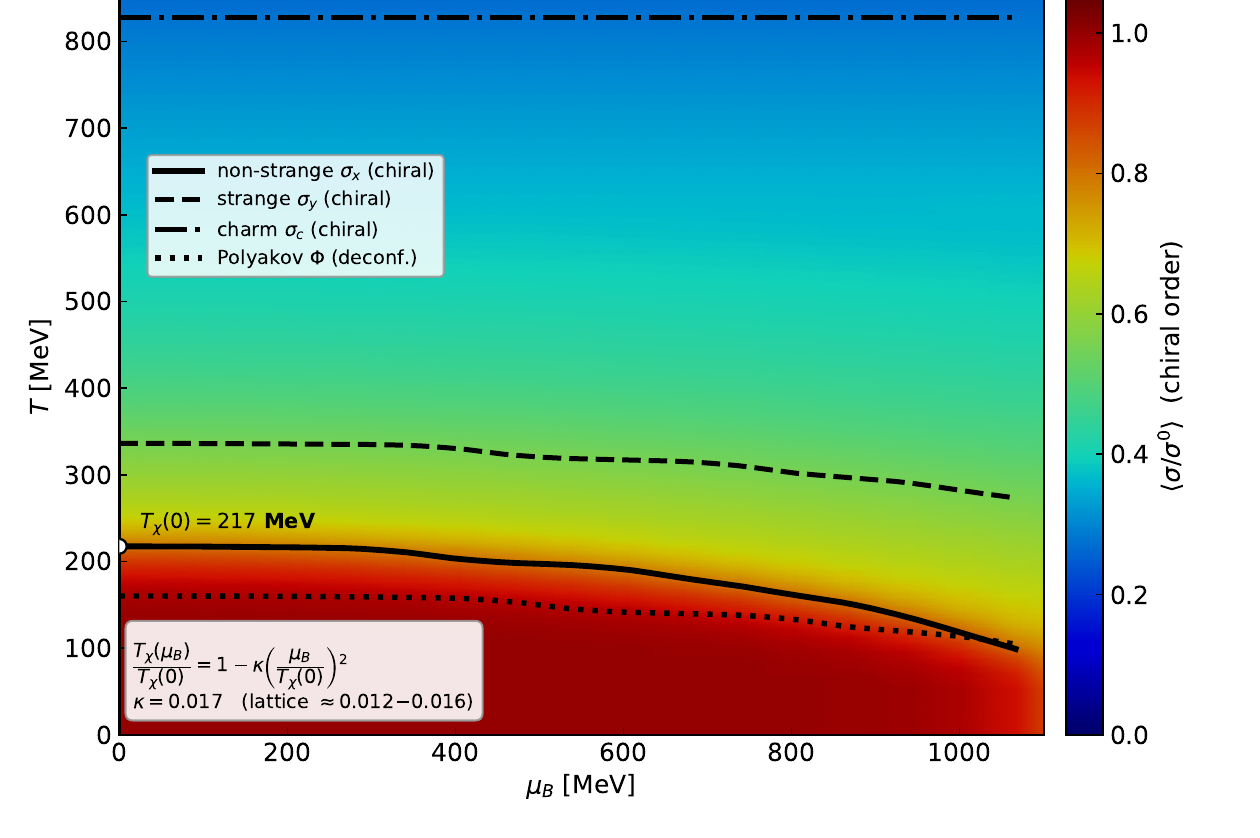}
\caption{Chiral and deconfinement phase structure in the $(T,\mu_B)$ plane for the $SU(3)$ (left) and $SU(4)$ (right) PLSM. The color scale is the combined chiral order parameter $\langle\sigma/\sigma^0\rangle$ (red means broken while blue refers to restoration). Solid, dashed, and dash-dotted black lines are the non-strange, strange and charm chiral pseudo-critical lines, respectively. The dotted line is the Polyakov deconfinement line. The inset gives the curvature of the chiral boundary, Eq.~(\ref{eq:kappa}), with $\kappa=0.018$ ($SU(3)$) and $0.017$ ($SU(4)$), compared with the lattice range.}
\label{fig:phase}
\end{figure}

Extending the mean-field solution to finite baryon chemical potential allows us to map the chiral and deconfinement pseudo-critical lines. Figure~\ref{fig:phase} collects the result for the two quark-flavors-configurations. The colored background is the combined chiral order parameter $\langle\sigma/\sigma^0\rangle$, which traces out the regions of broken (red) and restored (blue) chiral symmetry. The pseudo-critical lines are superimposed of each order parameter. The non-strange chiral crossover $T_\chi(\mu_B)$ bends downward with increasing density, as expected from the enhancement of chiral restoration by the baryon density, while the strange and for $SU(4)$ the charm chiral lines, together with the Polyakov deconfinement line, trace the sequential melting of the successive flavor sectors.

Near the origin the chiral boundary is well described by the standard curvature expansion
\begin{equation}
\frac{T_\chi(\mu_B)}{T_\chi(0)}=1-\kappa\left(\frac{\mu_B}{T_\chi(0)}\right)^2
+\mathcal{O}(\mu_B^4).
\label{eq:kappa}
\end{equation}
A statistical fit in the region $\mu_B\le400$~MeV gives a curvature $\kappa=0.018$ for $SU(3)$ and $\kappa=0.017$ for $SU(4)$. These values are close to, and slightly above, the continuum lattice range $\kappa\simeq0.012$--$0.016$~\cite{Bonati2015,Bellwied2015,
Bazavov2019,Cea2016}, a level of agreement that is typical of mean-field chiral models, which tend to somewhat overestimate the curvature. The near-equality of the two curvatures shows that the charm quark, which are inactive in the crossover region, seems to essentially have no influence on the shape of the light chiral boundary. Its effect on $T_\chi$ is limited to the small constant upward shift of about eight MeV already seen at $\mu_B=0$, induced by the weak charm back-reaction on the light condensate through the quark loop.

The most distinctive feature of the four-flavor phase diagram is the explicit appearance of three widely separated chiral scales. The non-strange line descends from $T_\chi(0)\simeq218$~MeV, the strange line sits near $330$~MeV at $\mu_B=0$, while the charm line lies around $820$~MeV, essentially independent of $\mu_B$ on the scale of the figure because the charm quark is too heavy to respond to the baryon density accessible here. This ordering is the finite-density image of the sequential restoration $u|d\to s\to c$ dictated by the quark-mass hierarchy. Up to roughly $\mu_B\simeq2\,T_\chi$, which we take as an edge of the mean-field domain, the transition stays a smooth crossover for every order parameter. The condensate drop does not steepen as $\mu_B$ grows. We find no sign of a first-order onset or of a critical endpoint. We therefore refrain from drawing a critical point, and restrict our focus on the displayed lines to the region where they are physically well defined. This absence of an endpoint is a statement about the mean-field model within its domain of validity, not about QCD simulations, where functional and lattice studies continue to probe the question~\cite{Fu2020,Gao2020,Fischer2019,Fodor2004}.

\section{Three-dimensional order-parameter surfaces}
\label{sec:3d}

\begin{figure}[htb]
\includegraphics[width=0.49\textwidth]{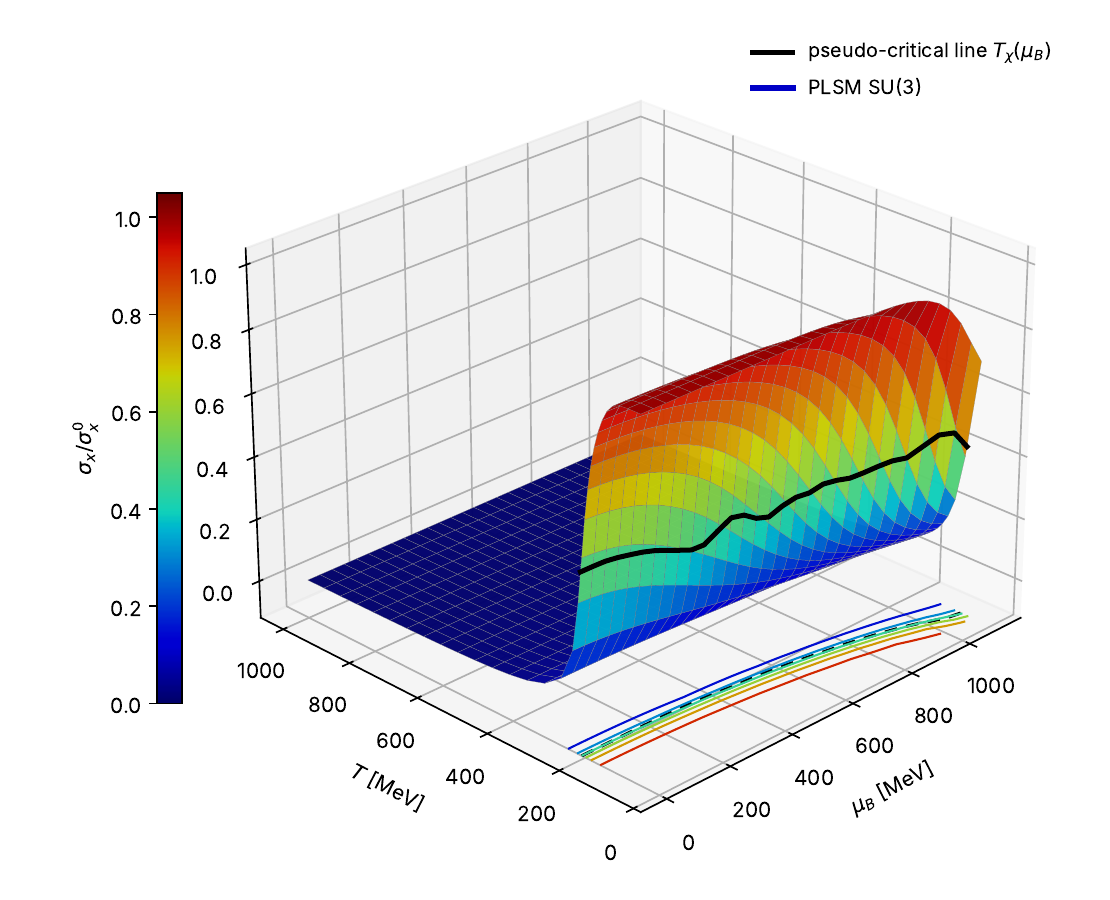}\hfill
\includegraphics[width=0.49\textwidth]{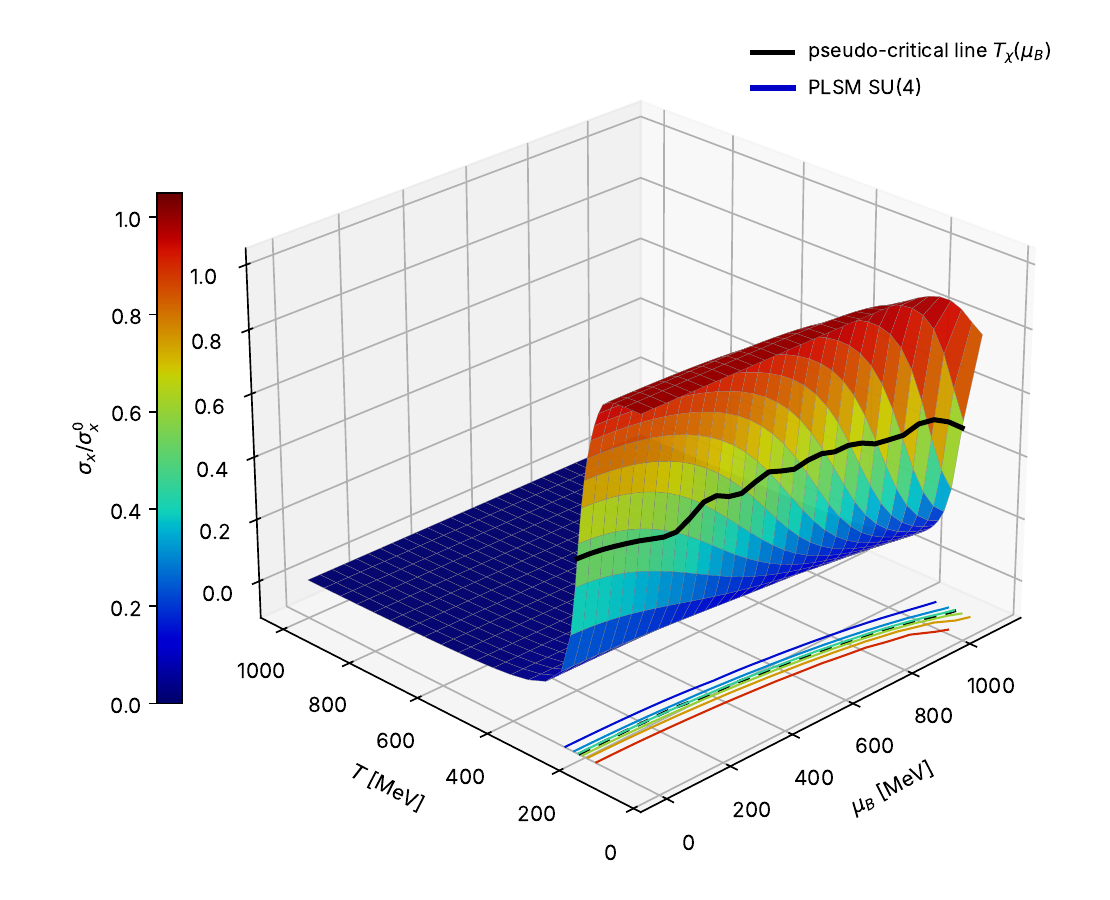}
\caption{Three-dimensional plot for $(T,\mu_B)$ surface of the normalized non-strange condensate $\sigma_x/\sigma_x^0$ for the $SU(3)$ (left) and $SU(4)$ (right) PLSM. The solid black curve is the chiral pseudo-critical line $T_\chi(\mu_B)$, pinned at $\mu_B=0$ to $210$~MeV ($SU(3)$) and to $218$~MeV ($SU(4)$).}
\label{fig:3dsx}
\end{figure}

\begin{figure}[htb]
\includegraphics[width=0.49\textwidth]{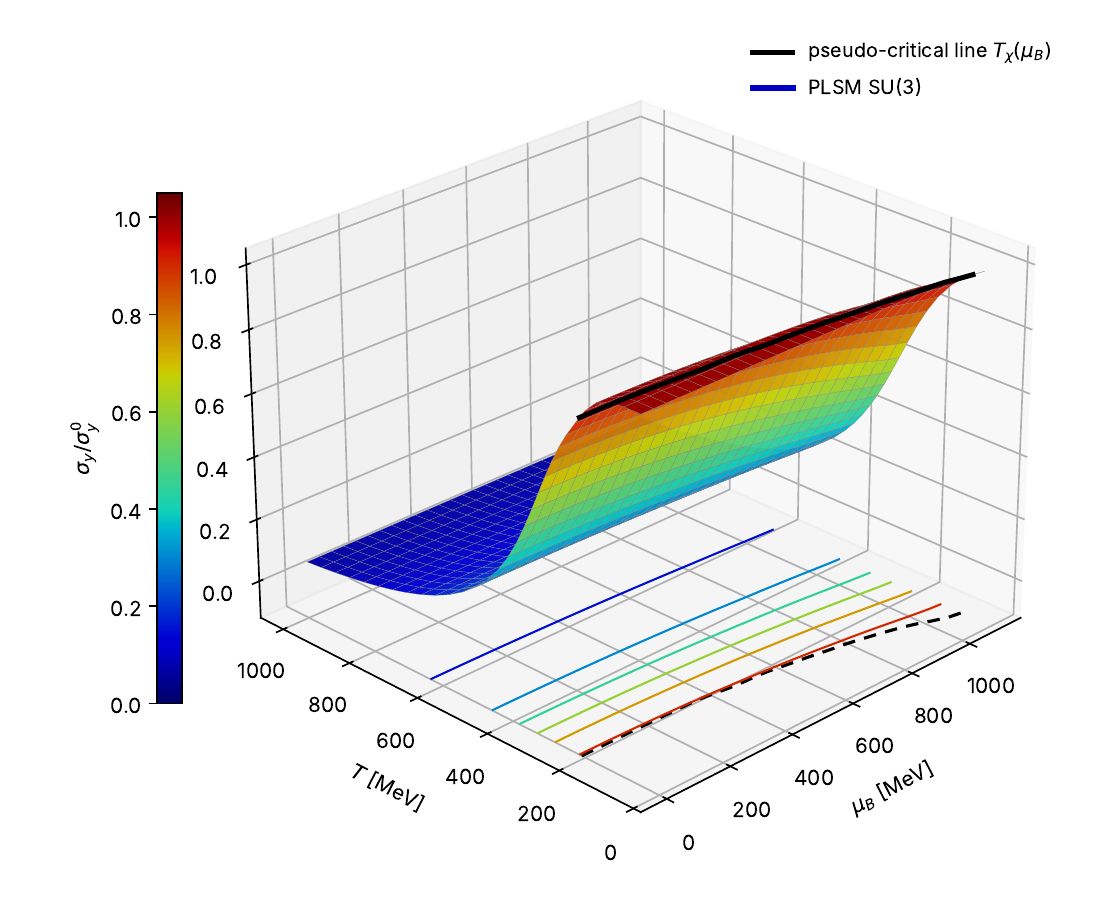}\hfill
\includegraphics[width=0.49\textwidth]{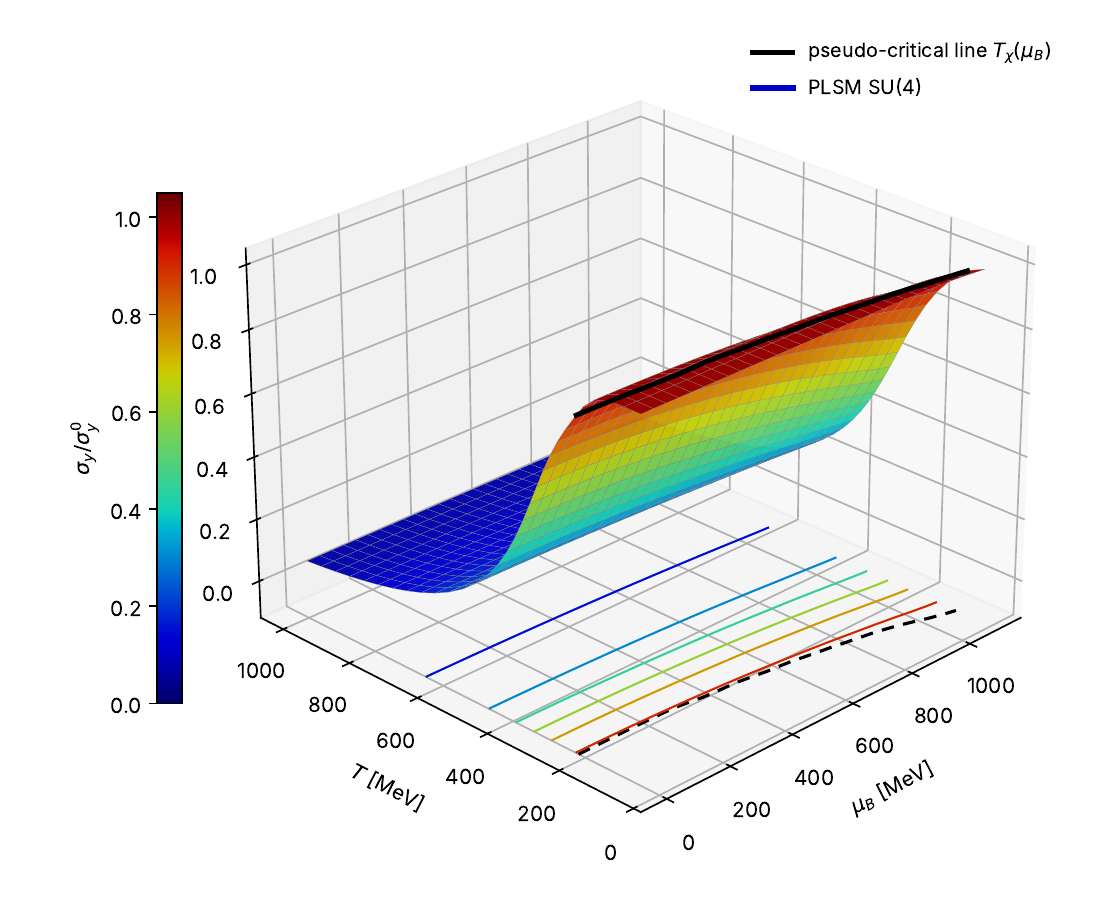}
\caption{The same as in Fig.~\ref{fig:3dsx} but here for the normalized strange condensate $\sigma_y/\sigma_y^0$ for the $SU(3)$ (left) and $SU(4)$ (right) PLSM. The strange condensate melts over a broader temperature range than the non-strange one and depends more weakly on $\mu_B$.}
\label{fig:3dsy}
\end{figure}

\begin{figure}[htb]
\centering
\includegraphics[width=0.49\textwidth]{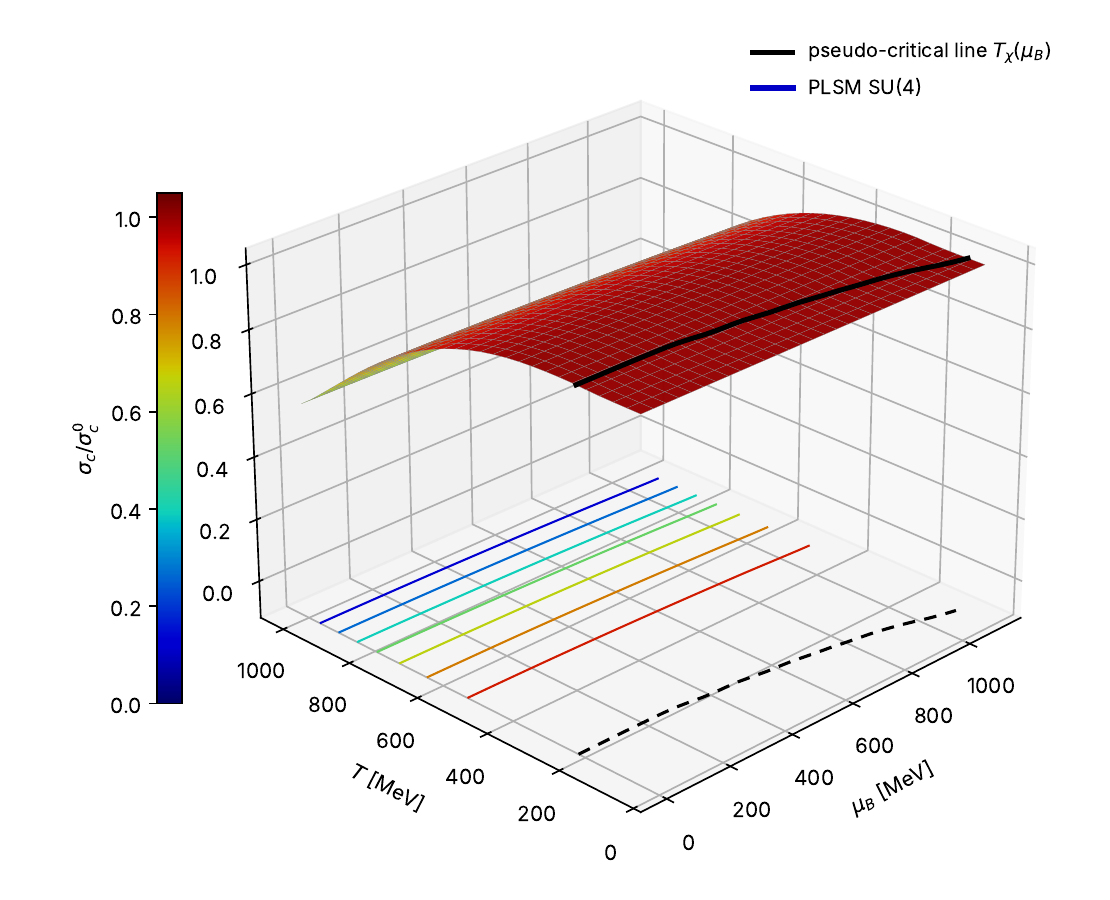}
\caption{The same as in Fig.~\ref{fig:3dsy} but here for the normalized charm condensate $\sigma_c/\sigma_c^0$ for the $SU(4)$ PLSM (the charm sector is absent in the $SU(3)$ model). The charm condensate stays close to its vacuum value over essentially the entire plane and is nearly independent of $\mu_B$.}
\label{fig:3dsc}
\end{figure}

\begin{figure}[t]
\includegraphics[width=0.49\textwidth]{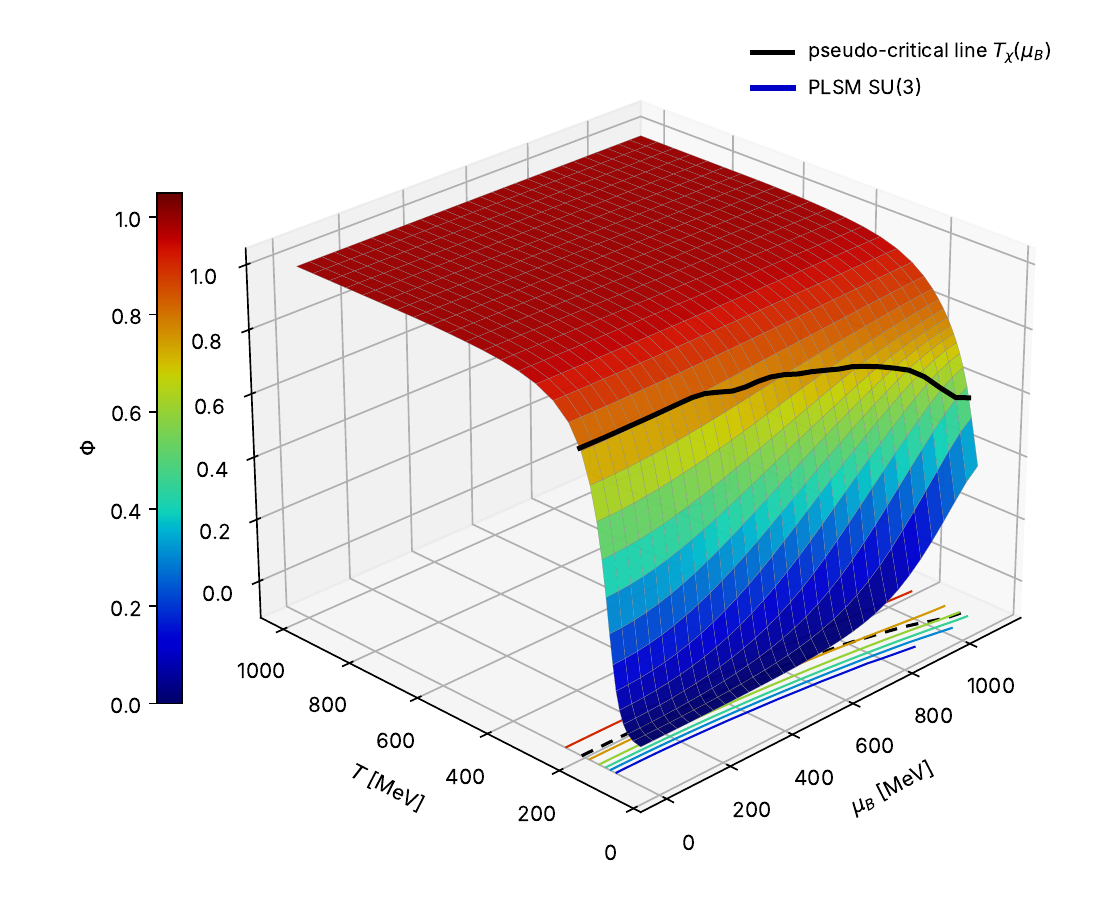}\hfill
\includegraphics[width=0.49\textwidth]{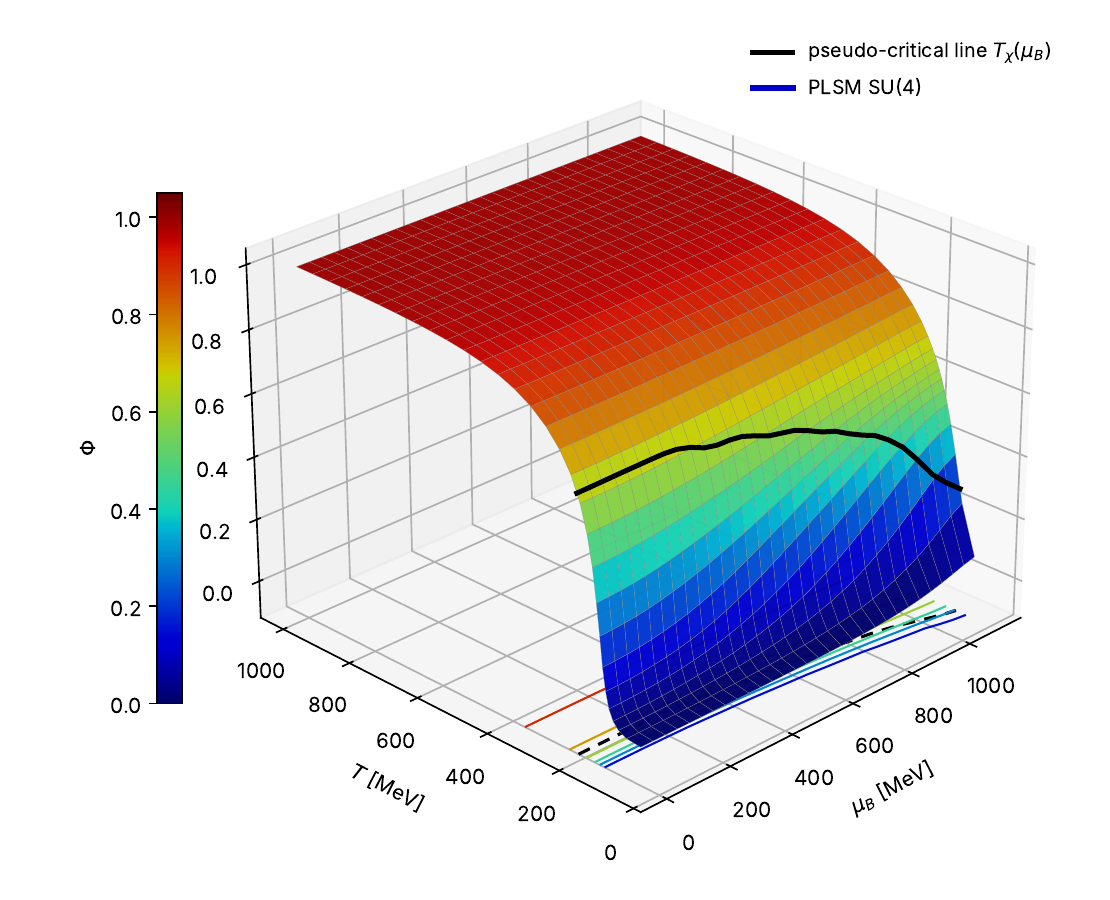}
\caption{Three-dimensional plot for $(T,\mu_B)$ surface of the Polyakov loop $\Phi$ for the $SU(3)$ (left) and $SU(4)$ (right) PLSM. The loop rises from the confined value $\Phi\simeq0$ to the deconfined plateau $\Phi\to1$, with a weak $\mu_B$ dependence. The solid black curve is the chiral pseudo-critical line $T_{\chi}(\mu_B)$.}
\label{fig:3dphi}
\end{figure}

The pseudo-critical lines of Section~\ref{sec:phase} are contours of a more complete object. It is shown the full dependence of each order parameter on the two thermodynamic variables $T$ and $\mu_B$. To make that dependence explicit, and to visualize how the sequential melting of the flavor sectors is organized across the whole plane, we display in Figs.~\ref{fig:3dsx}--\ref{fig:3dphi} the three-dimensional surfaces of the normalized condensates and of the Polyakov loop over the $(T,\mu_B)$ domain. In every panel the solid black curve riding on the surface is the chiral pseudo-critical line $T_\chi(\mu_B)$ of Eq.~(\ref{eq:kappa}). It is pinned at $\mu_B=0$ to the one-dimensional value of Tab.~\ref{tab:param} ($210$~MeV for $SU(3)$, $218$~MeV for $SU(4)$), so that the surfaces, the one-dimensional condensate curves of Fig.~\ref{fig:cond} and the EoS figure all mark one and the same transition temperature. The floor projection reproduces the two-dimensional phase boundary. The colored contours on the base are iso-level lines of the plotted field.

The non-strange condensate $\sigma_x/\sigma_x^0$, Fig.~\ref{fig:3dsx}, forms a single smooth step that descends from near unity in the low-$T$, low-$\mu_B$ corner to essentially zero in the deconfined region. The step does not sharpen into a discontinuity anywhere on the surface. Along every $\mu_B$ ray the drop retains the same crossover width, which is the geometric statement of the absence of a first-order line or critical endpoint within the mean-field domain. The ridge of steepest descent, traced by the pseudo-critical line, bends gently toward lower temperature as $\mu_B$ increases, reproducing the downward curvature quantified by $\kappa$ in Section~\ref{sec:phase}. We find that the $SU(3)$ and $SU(4)$ surfaces for $\sigma_x$ are nearly indistinguishable in shape, confirming at the level of the full surface that the charm sector does not reshape the light chiral transition. The only visible difference is the overall position of the step, shifted upward by the few-MeV charm back-reaction.

The strange condensate $\sigma_y/\sigma_y^0$, Fig.~\ref{fig:3dsy}, seems to melt over a much broader temperature interval. Its surface is a gentle slope rather than a sharp step. At any fixed $\mu_B$ a substantial fraction of the strange condensate survives well above the non-strange crossover. The half-melting contour lies tens of MeV above the $\sigma_x$ ridge. The $\mu_B$ dependence is correspondingly milder, because the larger strange constituent mass makes the strange sector less responsive to the baryon density than the light sector. The strange ridge bends down more slowly than the non-strange one. Here too the three- and four-flavor surfaces coincide, as they must, since the strange sector is common to both models and the charm back-reaction on it is negligible.

For the four-flavor model the charm condensate $\sigma_c/\sigma_c^0$, Fig.~\ref{fig:3dsc}, is the most striking surface precisely because it is almost featureless over the entire region plotted. It remains pinned near its vacuum value across the whole $(T,\mu_B)$ plane that the light and strange sectors traverse, and only in the far high-temperature corner does it begin a slow descent. It is effectively flat in $\mu_B$, reflecting the inability of the baryon chemical potential (density) accessible here to excite so heavy a flavor. The charm surface is thus a direct visual confirmation that the charm chiral scale is decoupled from the light dynamics. It provides the inert backdrop against which the sequential ordering $u|d\to s\to c$ is read off. As one raises the temperature at fixed $\mu_B$, the three surfaces collapse one after another, in order of increasing constituent mass.

Finally, the Polyakov loop $\Phi$, Fig.~\ref{fig:3dphi}, shows the complementary, deconfinement face of the transition. Its surface rises from the confined value $\Phi\simeq0$ at low temperature to the deconfined plateau $\Phi\to1$. With the steep rise located just below the chiral ridge, it is consistent with the one-dimensional ordering of Fig.~\ref{fig:cond} in which statistical deconfinement sets in marginally before chiral restoration. The $\mu_B$ dependence of the Polyakov surface is weak, since the loop responds to the deconfinement of color rather than directly to the baryon density. Its gentle tilt toward lower $T$ with increasing $\mu_B$ mirrors the downward bending of the chiral boundary and keeps the two crossovers locked together across the plane. The $SU(3)$ and $SU(4)$ Polyakov surfaces are practically identical, again because the heavy charm quark adds no thermally active color charge in this window.

Taken together, the four surfaces give a compact, global picture of the chiral and deconfinement dynamics. A single light step locked to the Polyakov rise, a broader strange slope above it, and a frozen charm plateau far above both. All bending coherently and gently toward lower temperature as the density grows. There is no steepening that would signal a critical endpoint anywhere the mean-field treatment can be trusted. We caution that the far corner of large $\mu_B$ and low $T$, where a critical endpoint is conjectured in the full theory, lies at the edge of the mean-field domain of validity. The surfaces are smooth there, but we refrain from reading quantitative structure from that region.

\section{Summary and conclusions}
\label{sec:conclusions}

We have carried out a systematic, observable-by-observable comparison of the three-flavor and four-flavor Polyakov linear-sigma models, with the explicit aim of isolating the thermal and phase-structural role of the charm quark. Working in the mean-field approximation, with the Polyakov sector of each model independently tuned to the continuum lattice EoS, we computed the light, strange, and charm chiral condensates, the Polyakov loop, and the full set of bulk thermodynamic observables. We also mapped out the chiral and deconfinement pseudo-critical lines in the temperature-density plane, both as two-dimensional boundaries and as complete three-dimensional order-parameter surfaces.

The two model configurations turn out to share the same crossover dynamics. In each case the non-strange condensate sets the chiral pseudo-critical temperature, $210$~MeV for $SU(3)$ and $218$~MeV for $SU(4)$. Both results position well above the lattice crossover, as expected from the mean-field approximation. The strange condensate melts more slowly, and the Polyakov loop deconfines in step with chiral restoration. Because the Polyakov sector was fixed to the lattice pressure and trace anomaly, the derived energy and entropy densities follow the HotQCD data over the whole overlap region in both models.

The charm quark, by contrast, plays almost no role in the transition itself. With $m_c$ far above $T_\chi$ the charm condensate hardly moves through the crossover, and the four-flavor EoS stays close to the three-flavor one, once each of them is referred to its own Stefan--Boltzmann limit until a small charm correction sets in above roughly $300$~MeV. That correction raises the ideal-gas d and lengthens the high-temperature tail, but it leaves the crossover region essentially unchanged. At finite density the picture is the same. The two chiral boundaries have nearly the same curvature, $\kappa\simeq0.017$--$0.018$, a little above the continuum lattice QCD value. The four-flavor phase diagram simply adds a third, well-separated chiral line for charm on top of that of light and strange ones, so that the ordering $u|d\to s\to c$ becomes visible directly --- both in the phase diagram and, globally, in the three-dimensional condensate surfaces. Over the whole range the mean-field solution can reach, the transition is a crossover. This is a statement about the model, not about QCD. It says only that no endpoint appears where the mean-field description is trustworthy. It should not be read as evidence either way for a critical point in the full theory.

With these limitations in mind, our results line up with the common expectation that charm decouples from the light QCD thermodynamics. They give a concrete sense of where, in temperature and baryon density, the decoupling starts to break down. Natural extensions include the conserved-charge fluctuations, which within the PLSM are sensitive probes of the deconfinement crossover~\cite{Sasaki2007,Skokov2010}. Thereby exploring both the transport properties of the medium and the full dynamics of the charm sector are possible. Moving beyond a simple spectator treatment reveals how charm quark actively probes the hot, dense matter generated in heavy-ion collisions.

\bibliographystyle{unsrtnat}
\bibliography{2026-30-30-Nourhane3rdPaper}

\end{document}